\documentstyle[PASJadd]{PASJ95}
\begin{document}
\setcounter{page}{1}

\title{ Polarization and Variations of BL Lacertae Objects}
\author{J.H. F{\bf{an}},$^{1,2}$ K.S. C{\bf{heng}},$^{1}$
L. Z{\bf{hang}}$^{1,3}$
\\[12pt]
$^1${ \it Department of Physics, The University of Hong Kong,
 Hong Kong, China}
\\
$^2${\it Center for Astrophysics, Guangzhou University, 
Guangzhou 510400, China}
\\
$^3${\it Department of Physics, Yunnan University, Kunming, China}
}

\abst{
 BL Lacertae objects are an extreme subclass of AGNs showing 
 rapid and large-amplitude variability, high and variable 
 polarization, and core-dominated radio emissions. If a 
 strong beaming effect is the cause of the extreme observation
 properties, one would expect that these properties would be 
 correlated with each other. 
 Based on the relativistic beaming model, relationships
 between the polarization and the magnitude variation in brightness,
 as well as the core-dominance parameter are derived and
 used statistically to compare with the observational data of 
 a BL Lacertae object sample.  
 The statistical results are consistent with these
 correlations, which suggests that the polarization, 
 the variation, and the core-dominance parameter are possible 
 indications of the beaming effect. }

\kword{
 BL Lacertae Objects: general -- galaxies: active--galaxies:jets
--galaxies:nuclei--galaxies: photometry--
Polarization}

\maketitle

\section{Introduction}

 BL Lacertae objects are generally described as a subclass of active
 galactic nuclei (AGNs), showing rapid and large-amplitude variation,
 variable and high polarization, a core-dominated non-thermal continuum
 with some BL Lacertae objects showing
  superluminal motion and high-energy  gamma-ray emissions ( 
 Angel, Stockman 1980; 
 Antonucci, Ulvestad 1985; 
 Fan et al. 1996; 
 Cheng et al. 2000;
 Ghisellini et al. 1993;
 Hartman et al. 1999; 
 Luna et al. 1993;
 Romero et al. 1995; 
 Shakhovsky,  Efimov 1999; 
 Stickel et al. 1993; 
 Takalo 1994; 
 Wills et al. 1992;
 Xie et al. 1994).    
 There are two subclasses of BL Lacertae objects, namely, the 
 radio-selected BL Lacertae objects (RBLs) and the X-ray selected BL 
 Lacertae objects (XBLs) from a survey or high-frequency peaked 
 BL Lacertae objects (HBLs) and low-frequency peaked
 BL Lacertae objects (LBLs) from the spectral energy distribution (SED).
 RBLs correspond to LBLs, while XBLs generally correspond to HBLs.

 The observational properties of RBLs are systematically different from
 those of XBLs.  The latter have a flatter spectral energy distribution 
 from the radio through X-ray regions, a higher starlight
 fraction, a higher observed peak of the emitted
 power from the radio through the X-ray 
 spectral energy distribution and convex  optical to X-ray continua. 
 Furthermore, 
 XBLs show lower polarization compared with RBLs, and they both occupy
 different regions in the effective spectral index plot (see 
 Morris et al. 1990;  Junnuzi et al. 1994; Giommi et al. 1995; 
 Sambruna et al. 1996; Fan, Xie 1996; Fan et al. 1997). 
 
 Observations show that the polarization in Mkn 421 is correlated
 with the brightness of the source (Tosti et al. 1998); a
 similar phenomenon was also
 observed from 3C 345 (Smith 1996).  Do the observations
 mean that there is a correlation between the polarization and the
 variations? In this paper, we discuss this correlation,
 and explain it in terms of a relativistic beaming effect. 
 In section 2, a relation between the polarization and  the variation
 is derived, and a comparison between the prediction and the
 observed data is presented.
 In section 3, we give some discussion and a conclusion.

\section{ Model}

 From the work of Padovani and Urry (1990) (see also Urry and Padovani,
 1995), the observed flux, 
 $S_{\rm{j}}^{\rm{\rm{ob}}}$, 
 of a relativistic jet is related to its intrinsic flux, 
 $S_{\rm{j}}^{\rm{\rm{in}}}$, 
 by $S_{\rm{\rm{j}}}^{\rm{\rm{ob}}} = \delta^{\rm{p}}S_{\rm{\rm{j}}}^{\rm{\rm{in}}}$, 
 where $\delta$,
 the Doppler factor of the jet, is defined by 
 $\delta = [\Gamma(1-\beta~cos ~\theta)]^{-1}$, $\beta$ is the velocity 
 in units of the speed of the light, $\Gamma = (1 - \beta^{2})^{-1/2}$ 
 is the Lorentz factor, and $\theta$ is the viewing angle.  The value 
 of $p$ depends on the shape  of the emitted spectrum and the detailed
 physics of the jet (Lind, Blandford 1985), $p = 3 + \alpha$ is for
 a moving sphere and $ p = 2 + \alpha$  is for the case of a continuous
 jet, where $\alpha$ is the spectral index.  However, because
 the emissions of an AGN can not be from the jet alone,  it
 is natural for one to suppose that they are from
 two components, namely the beamed and the unbeamed ones. 
 In this model, the  observed total flux, $S^{\rm{ob}}$, is the sum of 
 unbeamed, $S_{\rm{unb}}$ and beamed, $S_{\rm{j}}^{\rm{ob}}$ parts.
 Following the work of Urry and Shafer 1984 (see also Urry, 
 Padovani 1995), we assume that the intrinsic flux of the jet,
 $S_{\rm{j}}^{\rm{in}}$, is some fixed fraction, $f$, 
 of the unbeamed flux, 
 $S_{\rm{unb}}$, i.e., $S_{\rm{j}}^{\rm{in}} = fS_{\rm{unb}}$; we have 
 $ S^{\rm{ob}} = (1 + f\delta^{p}) S_{\rm{unb}}$.  For the case that
 $f$ is not a constant, its effect on our results is discussed
 in section 3. Furthermore, the flux is not totally
 polarized in the jet, and it is not unreasonable to assume that the jet
 flux consists of two parts, namely, the polarized and  unpolarized parts,
 with the two parts being proportional to each other, i.e., 
 $S_{\rm{j}}^{\rm{in}} = S_{\rm{jp}} + S_{\rm{jup}}$, 
 $S_{\rm{jp}} = \eta S_{\rm{jup}}$, where $\eta$
 is a coefficient which determines the polarization of the emission 
 in the jet.  The observed optical polarization can thus be expressed as 
\begin{equation}
P^{\rm{ob}} = {\frac{(1 + f) \delta_{\rm{o}}^{p}}{1 + f\delta_{\rm{o}}^{p}}} P^{\rm{in}}
\end{equation}
where the intrinsic polarization $P^{\rm{in}}$, is defined by
\begin{equation}
P^{\rm{in}} = {\frac{f}{1 + f}}{\frac{\eta}{1 + \eta}}
\end{equation}
and $\delta_{\rm{o}}$ is the optical Doppler factor (Fan et al. 1997).

\subsection{ Correlation between the Polarization and the Variation}

Using $S^{\rm{ob}}~=~(1+f\delta^{p})S_{\rm{unb}} \equiv S_{0}10^{-0.4m^{\rm{ob}}}$ 
and  relation (1), we can obtain 
\begin{equation}
{\frac{10^{-0.4m_{1}^{\rm{ob}}}P_{1}^{\rm{ob}}}{\delta_{1}^{3+\alpha}}}
={\frac{10^{-0.4m_{2}^{\rm{ob}}}P_{2}^{\rm{ob}}}{\delta_{2}^{3+\alpha}}} = k
\end{equation}
from which follows
\begin{equation}
{\frac{P^{\rm{ob}}_{1}}{P^{\rm{ob}}_{2}}} =
{\frac{\delta_{1}^{3+\alpha}}{\delta_{2}^{3+\alpha}}}~10^{0.4(m^{\rm{ob}}_{1}-m^{\rm{ob}}_{2})}
\end{equation}
 for the observations of any two epochs, 
 where $P^{\rm{ob}}$ and $m^{\rm{ob}}$ are
 the observed polarization and  magnitude, respectively,
 while $k$ is a constant of proportionality.
 Using
 $S^{\rm{ob}} = \delta^{3+\alpha}S^{\rm{in}} + S_{\rm{unb}}$,
 the ratio $\zeta = ({\frac{\delta_{1}}{\delta_{2}}})^{3+\alpha}$
 can be expressed as
$$ \zeta =
{\frac{S^{\rm{ob}}_{1} - S_{\rm{unb}}}{S^{\rm{ob}}_{2} - S_{\rm{unb}}}} >
{\frac{S^{\rm{ob}}_{1}}{S^{\rm{ob}}_{2}}}$$
 if $S^{\rm{ob}}_{1}$ is greater than $S^{\rm{ob}}_{2}$.
 In this sense, the ratio $\zeta$ can be written in the form
  $\zeta = ({\frac{S^{\rm{ob}}_{1}}{S^{\rm{ob}}_{2}}})^{\lambda}$, where
 the parameter $\lambda$ can be expected to be near unity, since 
 $\delta^{3+\alpha}S^{\rm{in}}$ is usually much greater than $S_{\rm{unb}}$.
 Thus, the ratio of polarizations (4) yields
\begin{equation}
{\frac{P_{1}}{P_{2}}} = 10^{0.4(\lambda - 1) \Delta m}\,
\end{equation}
where $\Delta m$ is the variation of magnitude, which follows
\begin{equation}
{\rm{log}}~ P_{1} = 0.4(\lambda - 1) \Delta m + {\rm{log}}~ P_{2} 
\end{equation}

 Relation (5) or (6), in principle, can be investigated based on the
 simultaneous
 observations of polarizations and  magnitudes (or flux density).
 Unfortunately, for most objects, there are no simultaneous 
 observations for  polarization and magnitude ( or flux densities).
 In this case,  if we set $P_{2}$ to be the minimum polarization, 
 $P_{\rm{min}}$,  we obtain
\begin{equation}
{\rm{log}}~ P (\%) = 0.4 (\lambda-1) \Delta m + {\rm{log}}~ P_{\rm{min}}.
\end{equation}

In order to avoid a possible observational bias (see section 3), 
we adopt 
the median polarization ($P_{\rm{Med}}$) as $P$ and the 
largest amplitude variation ($\Delta m_{\rm{Max}}$) as the variation 
$\Delta m$ in our discussion. 
The half value of the sum of the maximum and 
minimum polarization is taken as the median polarization ($P_{\rm{Med}}$).

\subsection{Results}

 From the available literature, we have compiled the corresponding 
 maximum and the minimum optical polarizations to obtain the median
 polarizations, largest amplitude variation, and the core-dominance
 parameters for 35 BL Lacertae  objects. They are listed in table 1, 
 in which
 Column 1 gives the name, 
 Columns 2 and 3 give the largest amplitude variation and the
         corresponding references;
 Columns 4 and 5 give the maximum and minimum optical
         polarization and the corresponding references;
 Columns 6 and 7 give the core-dominance parameter and the corresponding
         references.  
 The core-dominance parameter,
 $R$, is defined as the ratio of the radio core flux density to the flux 
 density in the extended lobes, and can be expressed as a function of 
 the Doppler factor ($\delta$) and $f$, i.e., $R~ =~ f\delta^{p}$ 
 (Ghisellini et al. 1993). 

\begin{table*}
\small
\begin{center}
Table~1. \hspace{4pt} Observation data of BL Lacertae Objects
\end{center}
\vspace{6pt}
\begin{tabular}{lcccccc}
\hline\hline\\
 Na me & $\Delta m_{\rm{Max}}$ &Ref & $P_{\rm{Max}}$ &Ref  & {\rm{log}}
$R$ & Ref \\
(1)     & (2)              & (3)  & (4)       & (5) & (6)     & (7) 
\\\hline
0048-097 & 2.7  &FL99  & 27.1/7.0 & AS80, CH84 & 0.97& W92\\
0118-272 & 1.05 &FL99  & 18.7     & SF97       &     &\\
0215+015 & 5.0  &P83   & 20.0/11.9& AS80, B86  &0.90 &W92\\
0219+428 & 2.97 &FL00B & 18.0/6.0 & FL99, RS85 &0.25 &W92\\
0235+164 & 5.3  &FL99  & 43.9/6.0 & ST93, FL96  & 2.25 &W92\\
0323+022 & 1.3  &F86   & 10.4/1.0 & J94        &      &\\
0521-365 & 1.4  &AS80  & 11.0/2.8  & B83, B86   & 0.01 &W92\\
0537-441 & 5.4  &FL00A & 18.8/10.1& IT90, IT88 & 2.3  &G93\\
0716+714 & 5.0  &F97   & 29.0/13.9& ST93, RS85 & 0.88 &T99\\
0735+178 & 4.6  &FL99  & 36.0/3.0 & IT90, FL96 & 3.4  &W92\\
0754+100 & 3.16 &FL00A & 26.0/4.0 & IT90, FL00A & 1.14 &W92\\
0818-128 & 3.78 &FL00A & 36.0/8.0 & FL99, FL00A  & 0.23 &W92\\
0823+033 & 1.41 &FL00A & 23.7/3.7 & IT88, KS90  &0.8  &G93\\
0828+493 & 2.0  &B90   & 7.9/1.4  & KS90        &         &\\
0829+046 & 3.58 &FL00A & 20.5/12.0& VW99, CH84 & 1.07 &W92\\
0851+202 & 6.0  &F98A  & 37.2/1.0 & ST93, AS80 & 3.5  &W92\\
1101+384 & 4.6  &FL99  & 16.0/0.0 & T98,  RS85  & 1.0  &G93\\
1144-397 & 1.92 &FL99  &  8.5/0.0 & IT88       &       &   \\
1147+245 & 0.99 &FL00A & 13.0/3.0 & IM82, IT88 & 1.42 &W92\\
1215+303 & 3.1  &FL99  & 14.0/8.0 & W78,  W92  & 0.27 &W92\\
1219+285 & 3.13 &FL00A & 20.0/2.0 & E99,  RS85 & 3.45 &W92\\
1308+326 & 4.17 &FL00A & 28.0/0.0 & ST93, MS81 & 1.70 &W92\\
1400+162 & 2.8  &Z81   & 14.0/4.0 & AS80, CH84 & 0.34 &ST\\
1418+546 & 4.8  &FL99  & 24.0/2.0 & FL99, AS80 & 1.77 &W92\\
1514-241 & 2.5  &FL00A&  8.0/2.0 & ST93, AS80 & 2.41 &W92\\
1519-273 & 2.43 &FL99  & 11.4/5.9 & IT90, IT88 & 0.9  &G93\\
1538+149 & 3.7  &FL99  & 32.8/5.0 & B86,  M90  & 0.95 &W92\\
1652+398 & 1.3  &FL99  &  7.0/2.0 & CH84, KS90 & 1.8  &G93\\
1727+502 & 2.1  &FL99  &  6.0/2.0 &       CH84 & 1.01 &W92\\
1749+096 & 2.7  &FL99  & 32.0/3.0 & B86,  KS90 & 2.83 &W92\\
1749+701 & 1.40 &B90   & 20.3/3.5 & W92,  RS85 & 1.10 &W92\\
1807+698 & 2.23 &FL00A & 12.0/0.0 & AS80, KS90 & 0.6  &G93\\
2155-304 & 1.85 &FL00B & 14.2/2.2 & PE96, B86  & 0.66 &W92\\
2200+420 & 5.31 &F98B  & 23.0/2.0 & ST93, AS80 & 2.41 &W92\\
2254+074 & 3.07 &FL00A & 21.0/8.8 & AS80, KS90 & 1.77 &W92\\
\hline
\end{tabular}

AS80: Angel, Stockman (1980);
B83: Bailey et al. (1983);
B86: Brindle et al. (1986);
B90: Bozyan et al. (1990);
CH84: Cruz-Gonzalez \& Huchra (1984);
E99: Efimov (1999);
F86: Feigelson et al. (1986);
F97: Fan et al. (1997);
F98A: Fan et al. (1998a);
F98B: Fan et al. (1998b);
FL96: Fan, Lin (1996);
FL99: Fan, Lin (1999);
FL00A: Fan, Lin (2000a);
FL00B: Fan, Lin (2000b);
G93: Ghisellini et al. (1993);
IM82: Impey et al. (1982);
IT88: Impey, Tapia (1988);
IT90: Impey, Tapia (1990);
J94: Jannuzi et al. (1994);
KS90: Kuhr, Schmidt (1990);
MS81: Moore, Stockman (1981);
M90: Mead et al. (1990);
P83: Pettini et al. (1983);
PE96: Pesce et al. (1996);
SF97: Scarpa, Falomo (1997);
ST: see the text;
ST93: Stickel et al. (1993);
T99: Tian et al. (1999);
T98: Tosti et al. (1998);
VW98: Visvanathan, Wills (1998);
W78: Wardle (1978);
W92: Wills et al. (1992);
Z81: Zekl et al. (1981)
\end{table*}

 For PKS 1219+285 (ON 231), which has been observed for about 100 years, 
 the early data observed by Wolf (1916) show that the object was as
 bright as 12 magnitude, which results in a  largest amplitude 
 variation of 5.4 mag. However, the 3 points discussed in the paper by
 Wolf (1916) are perhaps not certain, since they deviate from other 
 observations by about 2.3 mag,  if we do not take the 3
 early points into account.  The largest variation is thus only 3.13 
 (Fan, Lin 2000a), which is adopted in the present work.   
 For 1400+162, Jiang et al. (1999) obtained the VLBI total (165 mJy)
 and core (114 mJy) fluxes at 5~GHz, suggesting a core-dominance
 parameter of $R~={\frac{114}{165-114}}~=~2.2$. 
 The relevant  points are shown in figure 1. 
\begin{figure}
\vbox to3.0in{\rule{0pt}{3.0in}}
\includegraphics{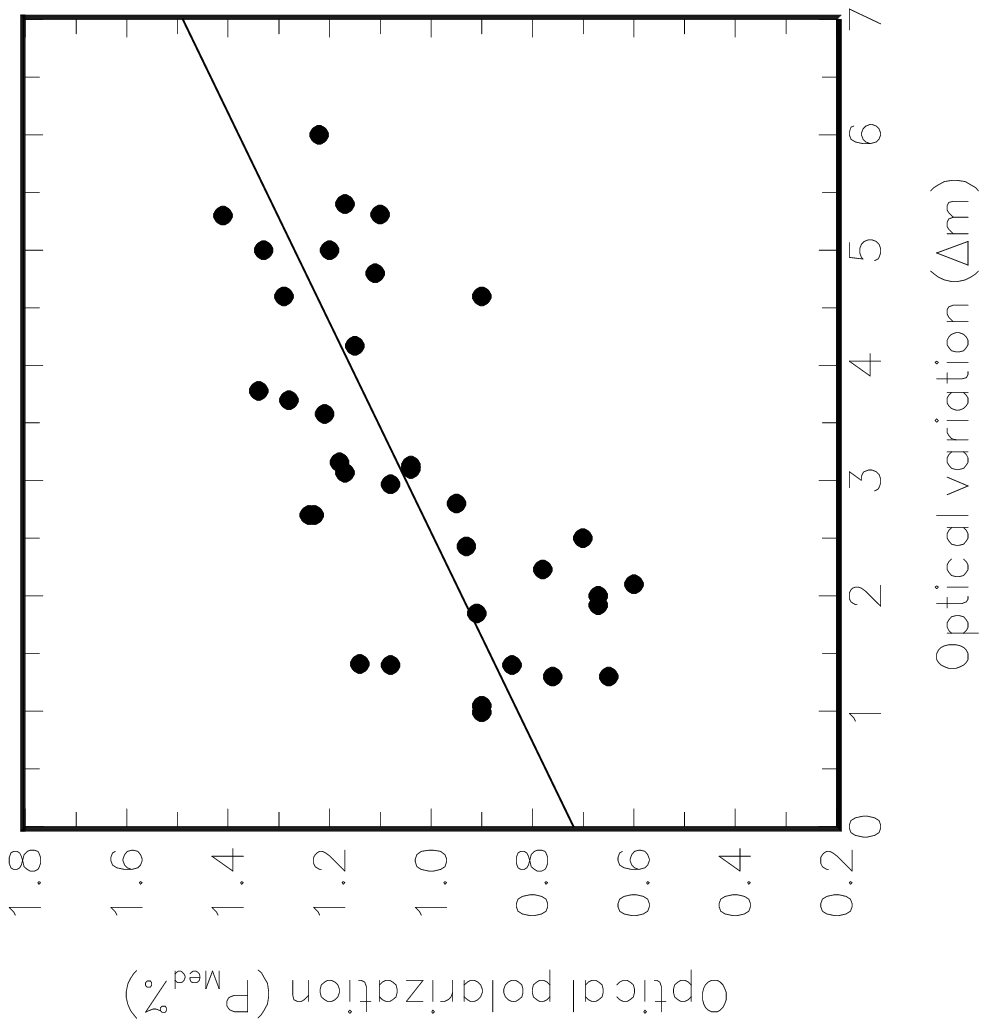}
\caption{Plot of polarization ( P$_{\rm{Med.}}$ (\%)) 
 against optical variation ($\Delta m$).}
\end{figure}
When a least-square regression fitting is  performed to the median
 polarization and the largest amplitude variation for these 35 objects, 
 the obtained result is
\begin{equation} 
 {\rm{log}}~ P(\%) = (0.10 \pm 0.01) \Delta m + 0.72 \pm 0.04
\end{equation}
 with a correlation coefficient of $r = 0.641$ and a chance 
 probability of $p = 6.32 \times 10^{-5}$.
 The best-fit result shown in figure 1 with a solid line
 implies that the parameters $\lambda = 1.25 \pm 0.12$
 and $P_{\rm{min}}$ = 5.2\%. 
 In addition, the relation of $\zeta$ gives  
  ${\rm{log}}~ R_{2} = 0.4 \lambda \Delta m + {\rm{log}}~ R_{1}$, 
 since  $R = f\delta^{3+\alpha}$, which can be tested by observations
 if we let $R_{2}~=~R_{\rm{Max}}$, $R_{1}~=~R_{\rm{min}}$, and 
 $\Delta m~=~\Delta m_{\rm{Max}}$. Therefore, 
 ${\rm{log}}~ R_{\rm{Max}}~=~0.4\lambda\Delta m_{\rm{Max}}~+~c_{1}$;
  here c1 is a constant
 associated with the minimum core-dominance, $R_{\rm{min}}$.
 The fitting result, $\lambda = 1.25 \pm 0.12$, predicts that 
 ${\rm{log}}~ R_{\rm{Max}} = (0.50 \pm 0.05 )\Delta m_{\rm{Max}} +c1$.
 When a linear regression was performed on the core-dominance
 parameter and the largest variation listed
 in table 1, a best-fit result, 
 ${\rm{log}} ~R = (0.31 \pm 0.11) \Delta m + 0.39 \pm 0.40 $,
  with a correlation
 coefficient $r = 0.45$ and a chance probability $p = 1\%$ was obtained.
 The best-fit result $0.31 \pm 0.11$ does not conflict with  
 the predicted slope $0.50\pm0.05$. 
 
\section{Discussion and Conclusion}

 BL Lacertae objects are characterized  by the observational properties
 mentioned in the introduction. 
 The beaming model was adopted to explain both 
 the particular observational properties and some observational
 differences between RBLs and XBLs (see Xie et al. 1991; Fan et al. 1993,
 1997; Fan, Xie 1996; Georganopoulos, Marscher 1999), although
 the viewing angle alone can not explain all of the difference between
 the two subclass BL Lacertae objects (Sambruna et al. 1996).  
 To discuss the intrinsic
 properties of BL Lacertae objects, one should know the boosting factor. 
 Many authors  have worked on this topic (Xie et al. 1991;
 Ghisellini et al. 1993; Lahteenmaki 1999; Fan et al. 1999). 
 Recently, using the  time variability, we determined the boosting 
 factor and other physical parameters of Blazars
 (Cheng, et al. 1999).

 It is believed that the particularly observed properties of
 BL Lacertae objects are associated with the beaming effect;
 if so, there should be a correlation between those properties. 
 In 1996, Smith reported that the polarization in 3C 345 is 
 strongly correlated with the brightness.
 Very recently, Tosti et al. (1998) also 
 found that there is a clear correlation between the variation of
 polarization degree and the brightness during the Mkn 421 1997
 outburst. A similar situation has also been found in 3C 66A 
(Efimov, Shakhoskoy 1998a)  and ON 231 (Efimov, Shakhoskoy 1998b).

 However, for most objects, the polarization and  magnitude variations 
 are not observed simultaneously. In addition,
 from an observational point of view, it is known that bright
 sources are observed frequently by photometry and polarimetry. In this
 sense, there is a tendency to detect a larger variation and higher 
 polarization in those brighter sources. This kind of bias would result
 in that those objects with a larger variation are also those with 
 a higher polarization. 
 To reduce this kind of bias, we have studied the median
 polarization of these objects. We chose the half value
 of the sum of the maximum and minimum polarization as the median
 polarization and used this value to investigate the correlation between
 the polarization and the variation for a sample of 35 BL Lacertae
 objects.  
 Nevertheless, we have also used the observed maximum polarization to
 analyze equation (7); the correlation coefficient is 0.65. Although this
 chance probability is better than that of using the median polarization 
 by a factor of 2, a possible observational bias may exist.
 
 It is known that XBLs are not as strongly beamed, and that the observed
 data can be taken as the intrinsic data to some extent (Fan, Xie 1996).
 In this sense,  the polarization of XBLs, which is  $ \sim5 \%$ on
 average (see Jannuzi et al. 1994; Fan 1999), can be taken as
 the minimum polarization of BL Lacertae objects. The best-fit
 result for the polarization and the variation implies a minimum 
 polarization of 5.2\%, which is consistent with the observation result
 of XBLs (Jannuzi et al. 1994).
 Our results indicate that the polarization is correlated with
 the variation, which is also consistent with the
 observation results in both 3C 345 and Mkn 421.

 The polarization is  found to be associated with the core-dominance
 parameter (see Wills et al. 1992 and reference therein)
 with a high polarization corresponding to a large {\rm{log}} $R$.
 From relation (3), one can obtain a relation between the polarization
 and the core-dominance parameter for a certain magnitude:
 $$P^{\rm{ob}} = k~\delta^{3+\alpha}~10^{0.4m^{\rm{ob}}}$$
\begin{equation} 
 =\left({\frac{k}{f}} \right)10^{0.4m^{\rm{ob}}}
(f\delta^{3+\alpha}) = c(m)R~\propto R
\end{equation}
 where $c(m)~=~(\large{{\frac{k}{f}}})10^{0.4m^{\rm{ob}}}$ 
 is a parameter that depends on the magnitude. The relation shows that the
 high polarization is associated with a large core-diminance parameter.

 From the adopted $c(m)$ = 0.01, 0.1, 1, 10, 
 several curves are obtained (figure 2);
  the curves fit the observation data
 well. The differences in $c(m)$ are from the differences in the 
 magnitudes amongst the objects.  From a catalogue by 
 Hewitt and Burbidge 
 (1993), the maximum  magnitude difference among the considered 
 objects is 5 mag,   which gives a difference of 100 in $c(m)$;
 this value does not conflict with the difference of $c(m)$ 
 adopted in the present paper, because BL Lacertae objects 
 are variable. The variability of the objects, themselves, should 
 result in a larger than 100 difference in $c(m)$. In addition, 
 although we used a fixed $f$ in our discussion, $f$ is likely 
 to be variable,
 which would result in a dilution of the correlation (cf. figure 2).
 Besides, the variation in $\eta$ would also affect the correlation.

\begin{figure}
\vbox to3.0in{\rule{0pt}{3.0in}}
\includegraphics{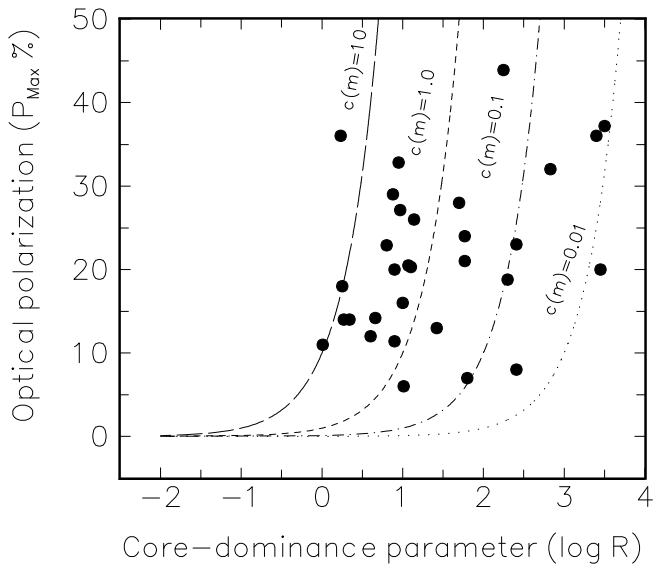}
\caption{
Plot of polarization ($P_{\rm{Max}} \%$) vs core-dominance parameter
({\rm{log}} R)}
\end{figure}

 We have tried to separate the data into three subgroups, 
 i.e., the lower ($\Delta m~<~ 2.0$ mag), intermediate 
 ( 2.0~mag $\leq~ \Delta m < 4.0$ mag), and higher 
 ($\Delta m ~\geq~ 4.0$ mag) amplitude variation groups, and
 found that the higher amplitude variation group shows a better linear
 relation between the polarization ($P$) and  the variation ($\Delta m$).
 We suspect that because the $P$ for the sources in this group 
 may be obtained when the sources are bright, they have similar
 magnitudes ($m$). However, the sample of such a group is small and 
 the statistical significance is very low. More observations
 in the bright state should confirm the result.

 In this work, the correlation between the polarization and the
 variation was derived and explained in the relativistic beaming
 frame for a BL Lacertae object sample. 
 Our statistical results show  that the particularly observed
 properties of BL Lacertae objects, such as the large-amplitude
 variation, core-dominance, and high polarization are associated with
 the relativistic beaming model. The mutual relationships are 
 consistent with the beaming model, suggesting that those observational 
 properties are  possible indications of a beaming effect. 

\par
\vspace{1pc} \par
 We are grateful to the referee for his/her suggestions, comments, and
 remarks. We thank Dr. B.J. Wills  for her 
 sending us a publication.
 This work is partially supported by the 
 Outerstanding Researcher Award of the University of Hong Kong, 
 the Croucher Senior Research Fellowship,
 the National Natural Science Foundation of China(19973001) and 
 the 973 Project of China (NKBRSF G19990754).

\section*{Reference}
\small

\re
Angel, J.R.P., \& Stockman, H.S.\ 1980, ARA\&A, 18, 321
\re
Antonucci, R.R.J., \& Ulvestad, J.S. \ 1985, ApJ, 294, 158
\re
Bailey, J., Hough, J.H., \& Axon, D.J.\ 1983, MNRAS, 203, 339
\re
Bozyan, E.P., Hemenway, P.D., \& Argue A.N.\
  1990, AJ, 99, 1421
\re
Brindle, C., Hough, J.H., Bailey, J.A., Axon, D.J., \& 
 Hyland, A.R.\ 1986, MNRAS, 221, 739 
\re
Cheng, K.S., Fan, J.H., \& Zhang, L.\ 1999, A\&A, 352, 32
\re
Cheng, K.S., Zhang, X., \& Zhang, L.\  2000, ApJ, 537, 80
\re
Cruz-Gonzalez, I., \& Huchra, J.P.\ 1984, AJ, 89,
441 
\re
Efimov, Y.\ 1999, Blazar Data News, N. 15
\re
Efimov, Y. \& Shakhovskoy, N.M.\ 1998a, OJ-94 annual meeting
 1997 "Multifrequency monitoring of Blazars" , Publ. Ossrv. Astro. Univ.
di. Perugia, 3, 24
\re
Efimov, Y. \& Shakhovskoy, N.M.\ 1998b, BLAZAR DATA, 1, No 3. 
\re
Fan, J.H.\ 1999, ASP Conf. Ser. 159, 57
\re
Fan, J.H., Xie, G.Z., \& Bacon, R.\ 1999, A\&AS, 136, 13
\re
Fan, J.H., \& Lin, R.G.\ 1999, ApJS, 121, 131
\re
Fan, J.H., \& Lin, R.G.\ 2000a, ApJ, 537, 101 
\re
Fan, J.H., \& Lin, R.G.\ 2000b, A\&A, 355, 880
\re
Fan, J.H., Adam, G., Xie, G.Z., Cao, S.L., Lin, R.G., Qin,
Y.P., Copin, Y., \& Bai,J.M.\ 1998a, A\&AS, 133, 217 
\re
Fan, J.H., Xie, G.Z., Pecontal, E., 
 Pecontal, A., \& Copin, Y.\ 1998b, ApJ, 507, 173
\re
Fan, J.H., Cheng, K.S., Zhang, L., Liu, C. H.\ 1997, A\&A,
    327, 947
\re
Fan, J.H., \& Lin R.G.\ 1996, Publ. Yunnan Obs., 65, 8
\re
Fan, J.H., Xie, G.Z., \& Wen, S.L.\ 1996, A\&AS, 116, 409
\re
Fan, J.H., \& Xie, G.Z.\ 1996, A\&A 306, 55
\re
Fan, J.H., Xie, G.Z., Li, J.J., Liu, J., Wen, S.L., Huang,
R.R., Tang, Z.M., \& Wang, Y.J.\ 1993, ApJ, 415, 113
\re
Feigelson, E.D., Bradt, H., McClintock, J., 
  Remilard, R., Urry, C.M., Tapia, S., Geldzahles, B., 
  Johnston, K. et al.\ 1986, ApJ, 302, 337
\re
Georganopoulos, M., \& Marscher, A.P.\
 1999, ASP Conf. 159, 359
\re
Ghisellini, G., Padovani, P., Celotti, A., \& Maraschi L. \
 1993, ApJ, 407, 65 
\re
Giommi, P., Ansari, S.A., Micol, A.\ 1995, A\&AS, 109, 267
\re
Hartman, R.C., Bertsch, D.L, Bloom, S.D., Chen, A.W.,
Deines-Jones, P., Esposito, J.A., Fichtel, C.E., Friedlanden, D.P.,
 et al.\ 1999, ApJS, 123, 79
\re
Hewitt, A., \& Burbidge, G.\ 1993, ApJS, 87, 451 
\re
Impey, C.D., Brand, P.W.J.L., Wolstencroft, R.D., 
  \& Williams, P.M.\ 1982, MNRAS, 200, 19 
\re
Impey, C.D., \& Tapia, S.\ 1988, ApJ, 333, 666 
\re
Impey, C.D., \& Tapia, S.\ 1990, ApJ, 354, 124
\re
Jannuzi, B.T., Smith, P.S., \& Elston, R.\ 1994, ApJ, 428, 130
\re
Jiang, D.R., Tian, W.M., Dallacasa, D., Nan, R.\  1999, New
Astron. Rev., 43, 703
\re
Kuhr, H., \& Schmidt, G.D.\ 1990, AJ, 99, 1 
\re
 Lahteenmaki, A.\ 1999, Ph.D. thesis, Metsahovi Radio Obs.
\re
 Lind, K.R. \& Blandford, R.D.\ 1985, ApJ, 295, 358 
\re
 Luna, H.G., Martinez, R.E., Combi, J.A., Romero, G.E.\ 1993,
A\&A, 269, 77
\re
Mead, A.R.G., Ballard, K.R., Brand, P.W.J.L., Hough, J.H.,
 Brindle, C., \& Bailey, J.A.\
 1990, A\&AS, 83, 183 
\re
Moore, R.L, \& Stockman, H.S.\ 1981, ApJ, 243, 60
\re
Morris, S.L., Stocke, J.T., Gioia, I.M., Schild, R.E.,
Wolten, A., Maccacaro, T., \& Ceca, R.D.\ 
 1991, ApJ, 380, 49 
\re
 Padovani, P., \& Urry, C.M.\ 1990, ApJ, 356, 75
\re
Pesce, J.E., Urry, C.M., Pian, E., Maraschi, L., Trves, A.,
 Grandi, P., Kellgaard, R.I., \& Marshall, H.\
1996, ASP Conf. Ser., 110, 423
\re
Pettini, M., Hunstead, R.W., Murdoch, H.S., \& Blades, J.C.\
1983, ApJ, 273, 436 
\re
Romero, G.E., Combi, J.A., Vucetich, H.\ 1995, Ap\&SS,
  225, 183
\re
Sambruna, R., Maraschi, L., Urry, C.M.\ 1996, ApJ, 463, 444
\re
Scarpa, R., \& Falomo, R.\ 1997, A\&A, 325, 109 
\re
 Shakhovsky, N.M. \& Efimov, Yu. S.\ 1999, in Gamov Memorial
Conference held in Odessa, Ukraine August 16-22, 1999 
\re
Smith, P.S.\ 1996, ASP Conf. Ser. 110, 135 
\re
Stickel, M., Fried, J.W., Kuhr, H.\ 1993, A\&AS, 98, 393 
\re
Takalo, L.O.\ 1994, Vistas in Astron., 38, 77 
\re
Tian, W.W., Krichbaum, P.T., Witzel, A. et al.\ 1999, A\&A, in
preparation 
\re
Tosti, G., Fiorucci, M., Luciani, M., Efimov, Yu. S.,
Shakhovsky, N.M., Valtaoja, E., Teraestanta, H., Sillanpaa, A.\ 
1998, A\&A, 339, 41 
\re
Urry, C.M., \& Padovani, P.\ 1995, PASP, 107, 803 
\re
Urry, C.M., \& Shafer, R.A.\ 1984, ApJ, 280, 569 
\re
Visvanathan, N., \& Wills, B.J.\ 1998, AJ, 116, 2119
\re
Wardle, J.F.C.\ 1978, Pitt. Conf. on BL Lacs, ed A. H.
 Wolfe, 39
\re
Wills, B.J., Wills, D., \& Breger, M., Antonucci, R.R.J., 
\& Barvainis, R.\ 1992, ApJ,
398, 454
\re
Wolf, M.\ 1916, Astron. Nachr., 202, 415 
\re
 Xie, G.Z., Liu, F.K., Liu, B.F., Lu, R.W., Li, K.H., \& Zhu,
Y.Y.\ 1991, A\&A, 249, 65 
\re
Xie, G.Z., Li, K.H., Zhang, Y.H., Liu, F. K., Fan, J.H.,
 \& Wang, J.C.\ 1994, A\&AS, 106, 361 
\re
Zekl, H., Klare, G., Appenzeller, I.\ 1981, A\&A, 103, 342

\end{document}